\documentclass[usenatbib]{mnras}
\usepackage[T1]{fontenc}
\usepackage[latin9]{inputenc}
\usepackage{array}
\usepackage{amsmath}
\usepackage{graphicx}
\usepackage{esint}
\usepackage[authoryear]{natbib}

\makeatletter

\providecommand{\tabularnewline}{\\}

\usepackage{array}

\providecommand{\tabularnewline}{\\}

\makeatother

\begin{document}
\title{Spatio-temporal non-localities in a solar-like mean-field dynamo}
\author{V.V. Pipin}

\maketitle
Institute of Solar-Terrestrial Physics, Russian Academy of Sciences,
Irkutsk, 664033, Russia
\begin{abstract}
The scale separation approximation, which is in the base of the solar
mean field dynamo models, can be hardly justified both by observations
and theoretical applications to astrophysical dynamos.{ The
general expression for the mean turbulent electromotive force can
be written in integral form with convolution of the turbulent effects
and mean magnetic field variations over scales of the turbulent flows
and global scales of the mean field dynamo. Following results of DNS,
which had been reported earlier, we take the Lorentzian form of the
integral convolution kernels as an experimental fact. }It allows us
to approximate the governing equation for the mean electromotive force
by the reaction--diffusion type equation. Solution of the eigenvalue
problem reveals a few curious properties of the dynamo model with
the nonlocal mean electromotive force. We find a decrease of the critical
dynamo instability threshold, and an increase the dynamo periods of
the unstable modes, as reported in earlier studies. Simultaneously,
the nonlocal model shows substantially lower growth rate of the unstable
dynamo modes in proximity of the critical threshold than the model
which employ the scale separation approximation. %Also, for the nonlocal model, we find a number of the different oscillating and steady dynamo modes can be excited in the close vicinity of threshold of the first unstable dynamo mode. 
We verify these findings using the nonlinear solar dynamo model. For
the supercritical regime, when the $\alpha$ effect magnitude is about
twice of the instability threshold, the model shows the Parker's dynamo
wave solutions with the wave propagating from the mid latitude at
the bottom of the convection zone toward the solar equator at the
surface. In the weakly nonlinear regime, when the $\alpha$ effect
magnitude is near the instability threshold, the interference of the
dynamo modes of different spatial localization shows the Grand activity
cycles of a period about 300 years. 
\end{abstract}

\section{Introduction}

Since \citet{Parker1955}, the standard scenario of the solar magnetic
cycle is based on the hydromagnetic dynamo, which includes the cyclic
transformation between toroidal and poloidal components of the large-scale
magnetic field of the Sun. \citet{SKR1966} put this idea on the theoretical
background, proposing the mean-field electrodynamics framework. It
was summarized in a number of textbooks, see, e.g, \citet{Moffatt1978,Parker1979,Krause1980}
.

The key theoretical ideas can be formulated as follows. Let us consider
the high conductive turbulent media and decompose the magnetic field
$\mathbf{B}$ and velocity field $\mathbf{U}$ on to mean and fluctuating
parts: $\mathbf{B}=\bar{\mathbf{B}}+\mathbf{b}$, $\mathbf{U}=\bar{\mathbf{U}}+\mathbf{u}$.
Here, we use the small letters for the fluctuating part of the fields
and capital letters with a bar above for the mean fields. Substitution
of these decomposition into induction equation and the averaging over
ensemble of the random fields give us the evolution equation for the
mean magnetic field, 
\begin{equation}
\partial_{t}\bar{\mathbf{B}}=\mathbf{\nabla}\times\left(\mathbf{\bar{\boldsymbol{\mathcal{E}}}+}\bar{\mathbf{U}}\times\bar{\mathbf{B}}\right),\label{eq:mfe}
\end{equation}
where the mean electromotive force, $\bar{\boldsymbol{\mathcal{E}}}$,
\begin{equation}
\bar{\boldsymbol{\mathcal{E}}}=\overline{\boldsymbol{u}\times\boldsymbol{b}}.
\end{equation}
It expresses the effects of the turbulence on the mean magnetic field
evolution. To calculate $\bar{\boldsymbol{\mathcal{E}}}$, we have
to solve the governing equations for the fluctuating velocity and
magnetic field. After using the scale separation approximation this
can be done analytically, e.g., with the double scale Fourier transform
\citep{Roberts1975} and the different assumptions about closure of
the correlation's chain (see, \citealp{Roberts1975,Kitchatinov1994,Kleeorin1996,Radler2007}).
Also the numerical estimation of $\bar{\boldsymbol{\mathcal{E}}}$
is possible with the test-field method (\citealp{Rheinhardt2010,Warnecke2018}).
In this case we avoid the closure assumptions completely. The general
structure of $\bar{\boldsymbol{\mathcal{E}}}$ can be guessed from
the properties of symmetries of transformations of the velocity magnetic
field and the assumption about scale separation, as well. Following
to \citet{Krause1980} we write it as the Taylor expansion about spatial
variations of the mean magnetic fields, 
\begin{equation}
\overline{\mathcal{E}}_{i}=\alpha_{ij}\overline{B}_{j}+\eta_{ijk}\partial_{k}\overline{B}_{j}+...,\label{mult}
\end{equation}
where $\alpha_{ij}$ is the ``pseudo-tensor'' which changes the
sign under reflection symmetry transformation. The $\alpha_{ij}$
can be further decomposed into sum of symmetric an the antisymmetric
parts. {The symmetric part of the $\alpha_{ij}$ stands for
the turbulent generation $\alpha$-effect. The antisymmetric part
of the tensor, which can be represented by vector, corresponds to
the turbulent pumping (e.g., \citealp{Krause1980})}. The antisymmetric
part of the third rank tensor $\eta_{ijk}$ represents the turbulent
eddy diffusivity.

The spatial and temporal scale separation, which is employed in the
Eq(\ref{mult}) can be hardly justified by the solar observations.
Indeed, the observations show the continuum spectrum of scales in
variations of the photospheric magnetic field variation in space \citep{Vidotto2016}.
The scale separation in time variations pronounces greater, having
two considerable peaks at the scale of the solar rotation period and
the second peak corresponds to the 11-th year solar cycle. Still,
it shows the continuum spectrum with inclination of 2/3 in between
of these peaks \citep{Frick2020}. Similar results were found from
observations of the magnetic activity in the solar type stars having
the external convective envelope. Moreover, the very active fast rotating
stars can demonstrate the continuum temporal spectrum of the magnetic
activity \citep{Stepanov2020MN}. Breaking of the scale separation
assumption can be easily seen in solutions of the mean-field solar
dynamo models, as well. Those solutions often show strong spatial
variations of the mean magnetic field near boundaries of the dynamo
domain \citep{Chatterjee2011,Brandenburg2018a,Pipin2019c}.

To account the strong variations of the mean magnetic field in space
and time we have to retain the higher order derivatives in expression
of $\overline{\boldsymbol{\mathcal{E}}}$. \citet{Raedler1976IAU}
and \citet{Raedler1980} suggested that the general conditions expression
of $\overline{\boldsymbol{\mathcal{E}}}$ should be written as a convolution
between an integral kernel and the mean field, e.g.,

\begin{equation}
\mathcal{E}_{i}=\hat{G}_{ij}*\overline{B}_{j},\label{eq:nlc}
\end{equation}
where the asterisk means a convolution in space and time. Similarly
to the Eq(\ref{mult}), we can split it into two pieces and write
\citep{Rheinhardt2012},

\begin{equation}
\mathcal{E}_{i}=\hat{\alpha}_{ij}\ast\overline{B}_{j}+\hat{\eta}_{ijk}*\partial_{k}\overline{B}_{j}\label{eq:nlc0}
\end{equation}

The direct numerical simulations (see, \citealt{Brandenburg2002,Rheinhardt2012,Kandu2022B,Elstner2020G})
showed that in the spectral space the kernel $\hat{G}$ is close to
a Lorentzian form, i.e., $\hat{G}\sim$$\left(1+\mathrm{i}\omega\tau+\ell^{2}k^{2}\right)^{-1}$,
here $\tau$ corresponds to the turbulent turnover time and $\ell$
characterizes the length scale on which non-locality becomes important.
In this paper, similar to \citet{Brandenburg2018a}, we accept the
hypothesis, $\hat{G}\sim$$\left(1+\mathrm{i}\omega\tau+\ell^{2}k^{2}\right)^{-1}$.
The Lorentzian form of the kernel $G$ results into the partial equation
for the mean electromotive force in parabolic form, 
\begin{eqnarray}
\left(1+\tau\frac{\partial}{\partial t}+a_{E}\eta_{T}\nabla^{2}\right)\overline{\boldsymbol{\mathcal{E}}} & = & \overline{\boldsymbol{\mathcal{E}}}^{(0)},\label{eq:nlc1}\\
\overline{\boldsymbol{\mathcal{E}}}^{(0)} & = & \alpha_{ij}\overline{B}_{j}+\eta_{ijk}\partial_{k}\overline{B}_{j}\label{eq:nlc2}
\end{eqnarray}
where, $a_{E}\approx0-1$ is the spatial non-locality parameter \citep{Rheinhardt2012},
the RHS of the Eq(\ref{eq:nlc2}) corresponds to the local expression
of the mean electromotive force obtained either numerically, e.g.,
by the test-field method or analytically using the SOCA (second order
correlation approximation), e.g., \citet{Roberts1975} and \citet{Kitchatinov1994},
or the different forms of the tau approximations (see, \citealt{Kleeorin1996,Radler2003,Pipin2008a})
or the other analytical methods. We have to note that unlike SOCA,
the tau approximation is also valid for the cases of the developed
turbulence characterized by the high Reynolds numbers. \citet{Radler2007}
gave the comprehensive discussion of the analytical approaches for
calculation of the mean electromotive force. {We consider the
$\tau$ relaxation term in the Eq.(\ref{eq:nlc1}), which was suggested
originally by \citet{Blackman2002a}, as a free parameter. We study
two cases. In one case we put $\tau=\tau_{c}$, where $\tau_{c}$
is the solar profile of the convective turnover time. Our study considers
a constant value of $\tau$, as well. In this case we choose $\tau$
from the range of $\tau_{c}$ profile.}

Our goal is to study effects of the spatio-temporal non-localities
in the solar type dynamos and to compare them with the reference dynamo
model that utilizes the scale separation approximation. We construct
the nonlocal model using the zero order approximation of the mean
electromotive force ($\overline{\boldsymbol{\mathcal{E}}}^{(0)}$
from the Eq.(\ref{eq:nlc2})) obtained with the minimal tau approximation
by \citet{Pipin2008a}. We used it in the reference dynamo model of
\citet{Pipin2019c}. Next section discusses the model formulation.
After that, we consider the results of the eigen value problem and
the nonlinear runs of the nonlocal dynamo model. We resume the main
findings in the last section of the paper.

\section{Dynamo model}

Similarly, to \citet{Brandenburg2018a} we solve the magnetic field
evolution using the mean field dynamo induction equation, the Eq.(\ref{eq:mfe})
and the evolution equation for the mean electromotive force in form
of the Eq.(\ref{eq:nlc1}). The mean-field is decomposed into a sum
of the poloidal and toroidal components, as follows:

\[
\mathbf{\overline{B}}=\hat{\mathbf{\phi}}B+\nabla\times\left(A\hat{\mathbf{\phi}}\right)\,,
\]
The zero order approximation of the mean electromotive force is calculated
using the scale separation assumption an the minimal $\tau$ approximation
(see, \citealt{Pipin2008a}) It reads, 
\begin{equation}
\overline{\mathcal{E}}_{i}^{(0)}=\left(\alpha_{ij}+\gamma_{ij}\right)\overline{B}_{j}-\eta_{ijk}\nabla_{j}\overline{B}_{k},\label{eq:Ea}
\end{equation}
here, $\alpha_{ij}$ describes the turbulent generation of the magnetic
field by helical motions (the $\alpha$-effect), $\gamma_{ij}$ describes
the turbulent pumping, and $\eta_{ijk}$ is the eddy magnetic diffusivity
tensor. We take their analytical expression from results of \citet{Pipin2008a}
(hereafter P08). The $\alpha$-effect tensor includes effects of the
magnetic helicity, i.e., 
\begin{eqnarray}
\alpha_{ij} & = & C_{\alpha}\psi_{\alpha}(\beta)\alpha_{ij}^{(H)}+\alpha_{ij}^{(M)}\psi_{\alpha}(\beta)\frac{\left\langle \mathbf{a}\cdot\mathbf{b}\right\rangle \tau_{c}}{4\pi\overline{\rho}\ell_{c}^{2}},\label{alp2d}
\end{eqnarray}
where the full expressions of the kinetic helicity tensor $\alpha_{ij}^{(H)}$
and the tensor $\alpha_{ij}^{(M)}${, which defines the magnetic
helicity contribution, }are given in P08 and also in \citet{Pipin2022}
(hereafter P22), $\mathbf{a}$ and $\mathbf{b}$ are the fluctuating
vector-potential and magnetic field, respectively. The radial profiles
of the $\alpha_{ij}^{(H)}$ and $\alpha_{ij}^{(M)}$ depend on the
mean density stratification, profile of the convective RMS velocity
$u_{c}$ and on the Coriolis number $\Omega^{*}=2\Omega_{0}\tau_{c}$,
where $\Omega_{0}$ is the angular velocity of the star and $\tau_{c}$
is the convective turnover time. In our model we assume that the convective
turnover time corresponds to the turbulent relaxation time of the
$\overline{\boldsymbol{\mathcal{E}}}$. The magnetic quenching function
$\psi_{\alpha}(\beta)$ depends on the parameter $\mathrm{\beta=\left|\overline{\mathbf{B}}\right|/\sqrt{4\pi\overline{\rho}u_{c}^{2}}}$.
Its expression, as well as analytical expressions for $\alpha_{ij}^{(H)}$
and $\alpha_{ij}^{(M)}$ are given in Pipin\citeyearpar{Pipin2008a,Pipin2022}.

The magnetic helicity density evolution is governed by the balance
equation for the total magnetic helicity, $\left\langle \chi\right\rangle ^{(\mathrm{tot})}=\left\langle \mathbf{a}\cdot\mathbf{b}\right\rangle +\overline{\mathbf{A}}\cdot\overline{\mathbf{B}}$,
(see, \citet{Hubbard2012,Pipin2013c,Brandenburg2018}): 
\begin{equation}
\left(\frac{\partial}{\partial t}+\boldsymbol{\overline{\mathbf{U}}\cdot\nabla}\right)\left\langle \chi\right\rangle ^{(\mathrm{tot})}=-\frac{\left\langle \mathbf{a}\cdot\mathbf{b}\right\rangle }{R_{m}\tau_{c}}+\mathbf{\nabla\cdot}\eta_{\chi}\mathbf{\nabla}\left\langle \mathbf{a}\cdot\mathbf{b}\right\rangle ,\label{eq:helcon}
\end{equation}
where, we use ${\displaystyle 2\eta\mathbf{\left\langle b\cdot j\right\rangle }=\frac{\left\langle \mathbf{a}\cdot\mathbf{b}\right\rangle }{R_{m}\tau_{c}}}$
\citep{Kleeorin1999}. The second term in the RHS defines the diffusive
flux of the small-scale magnetic helicity density, we put $\eta_{\chi}=\frac{1}{10}\eta_{T}$
\citep{Mitra2010}; $R_{m}$ is the magnetic Reynolds number, we employ
$R_{m}=10^{6}$.

The anisotropic diffusion tensor was given by \citet{Pipin2008a}
: 
\begin{eqnarray}
\eta_{ijk} & = & 3\eta_{T}\left\{ \left(2f_{1}^{(a)}-f_{2}^{(d)}\right)\varepsilon_{ijk}+2f_{1}^{(a)}\frac{\Omega_{i}\Omega_{n}}{\Omega^{2}}\varepsilon_{jnk}\right\} \label{eq:diff}
\end{eqnarray}
where 
\begin{eqnarray*}
f_{1}^{(a)} & = & \frac{1}{4\Omega^{*\,2}}\left(\left(\Omega^{*\,2}+3\right)\frac{\arctan\Omega^{*}}{\Omega^{*}}-3\right),\\
f_{2}^{(d)} & = & \frac{1}{\Omega^{*\,2}}\left(\frac{\arctan\left(\Omega^{*}\right)}{\Omega^{*}}-1\right)
\end{eqnarray*}
The antisymmetric tensor $\gamma_{ij}$ stands for the turbulent pumping,
which is usually considered \citep{Krivodubskij1987,Warnecke2018}
as important ingredient of the solar dynamo process. Following \citet{Pipin2022},
we model it as follows, 
\begin{eqnarray}
\gamma_{ij} & = & \gamma_{ij}^{(\Lambda\rho)}+\frac{\alpha_{\mathrm{MLT}}u_{c}}{\gamma}\mathcal{H}\left(\beta\right)\mathrm{\hat{r}_{n}\varepsilon_{inj}},\label{eq:pump0}\\
\gamma_{ij}^{(\Lambda\rho)} & \!\!=\!\! & \!3\nu_{T}f_{1}^{(a)}\!\!\!\left\{ \!\!\left(\mathbf{\boldsymbol{\Omega}}\cdot\boldsymbol{\Lambda}^{(\rho)}\!\right)\!\!\frac{\Omega_{n}}{\Omega^{2}}\varepsilon_{\mathrm{inj}}\!-\!\frac{\Omega_{j}}{\Omega^{2}}\mathrm{\varepsilon_{inm}\Omega_{n}\Lambda_{m}^{(\rho)}}\!\!\right\} \label{eq:pump1}
\end{eqnarray}
where $\mathbf{\boldsymbol{\Lambda}}^{(\rho)}=\boldsymbol{\nabla}\log\overline{\rho}$,
$\mathrm{\alpha_{MLT}}=1.9$ is the mixing-length theory parameter,
$\gamma$ is the adiabatic exponent, $u_{c}$ is the RMS convective
velocity. The magnetic quenching function $\mathcal{H}\left(\beta\right)$
are given in the above cited paper (also, see, \citealt{Ruediger1995}).

We calculate the turbulent parameters using the mixing-length approximation
and the profile of the mean entropy. 
\begin{equation}
\mathrm{u_{c}=\frac{\ell_{c}}{2}\sqrt{-\frac{g}{2c_{p}}\frac{\partial\overline{s}}{\partial r}},}\label{eq:uc}
\end{equation}
where $\ell_{c}=\alpha_{\mathrm{MLT}}H_{p}$ is the mixing length,
$\alpha_{\mathrm{MLT}}=1.9$ is the mixing length parameter, and $H_{p}$
is the pressure height scale. The entropy profile is defined by solving
the mean-field heat transport equation (see, e.g., P22) for the rotating
convection zone. It deals with deviations of the mean entropy from
the reference state due to effect of rotation and the heat energy
sink and gain from evolution of the large-scale velocity and magnetic
field. To calculate the reference profiles of mean thermodynamic parameters,
such as entropy, density, temperature and the convective turnover
time, $\tau_{c}$, we use the MESA model \citep{Paxton2011,Paxton2013}.
The Eq.(\ref{eq:uc}) defines the profiles of the eddy heat conductivity,
$\chi_{T}$, eddy viscosity, $\nu_{T}$, and eddy diffusivity, $\eta_{T}$,
as follows, 
\begin{eqnarray}
\chi_{T} & = & \frac{\ell^{2}}{6}\sqrt{-\frac{g}{2c_{p}}\frac{\partial\overline{s}}{\partial r}},\label{eq:ch}\\
\nu_{T} & = & \mathrm{Pr}_{T}\chi_{T},\label{eq:nu}\\
\eta_{T} & = & \mathrm{Pm_{T}\nu_{T}},\label{eq:et}\\
\eta_{\mathcal{E}} & = & a_{E}\chi_{T}.\label{eq:etE}
\end{eqnarray}
{Here, we parameterize the $\eta_{\mathcal{E}}$ profile with
$a_{E}=0.1-1$, in following to \citep{Rheinhardt2012}}. The angular
velocity profile, $\Omega\left(r,\theta\right)$, and the meridional
circulation, $\mathbf{\overline{U}}^{(m)}$, are defined by conservation
of the angular momentum and azimuthal vorticity $\overline{\omega}=\left(\mathbf{\nabla}\times\mathbf{\overline{U}}^{(m)}\right)$,
\citep{Pipin2019c,Pipin2022}. In this paper we use the kinematic
models excluding effects of the $\mathbf{\overline{B}}$ on the large-scale
flow and heat transport. The model shows an agreement of the angular
velocity profile with helioseismology results for $\mathrm{Pr}_{T}=3/4$.
The dynamo models with local $\boldsymbol{\overline{\mathcal{E}}}$
show cycle period of $22$ years when $\mathrm{Pm}_{T}=10$ and $C_{\alpha}=0.04$2.
The level $C_{\alpha}$ is slightly above the critical threshold,
see the results in the next subsection.

\begin{figure*}
\centering \includegraphics[width=0.9\textwidth]{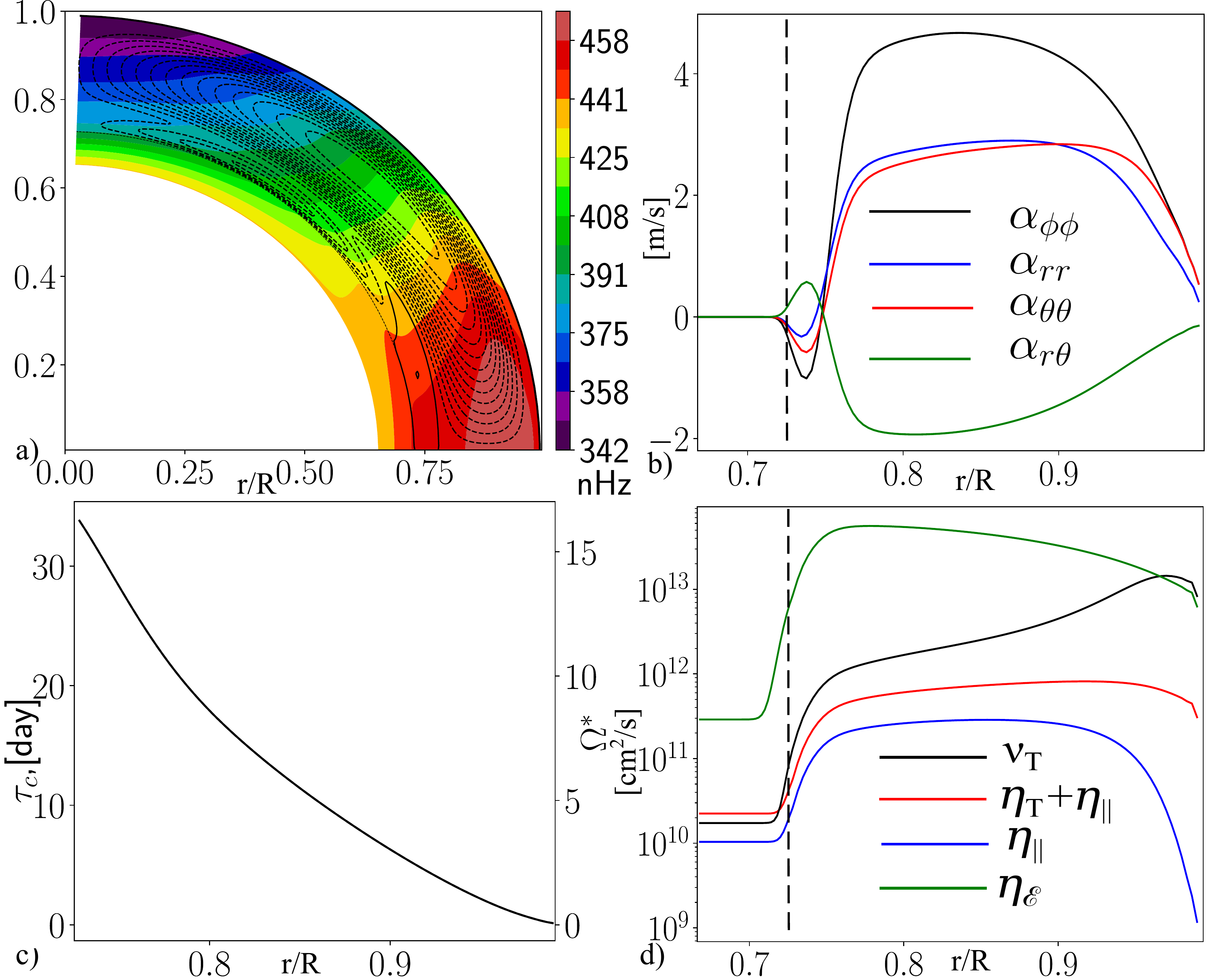} \caption{\label{fig1} a) The meridional circulation (streamlines) and the
angular velocity distributions; the magnitude of circulation velocity
is of 13 m/s on the surface at the latitude of 45$^{\circ}$; b) the
$\alpha$-effect tensor distributions at the latitude of 45$^{\circ}$,
the dash line shows the convection zone boundary; {c) radial
profiles of the convective turnover time, $\tau_{c}$, ( left y axis)
and the Coriolis number, $\Omega^{*}$, (right y axis);} d) radial
dependencies of the total, $\eta_{T}+\eta_{||}$, and the rotational
induced part, $\eta_{||}$, of the eddy magnetic diffusivity, the
eddy viscosity profile, $\nu_{T}$ and the the $\overline{\mathcal{E}}$
diffusivity profile for $a_{E}=1$; hereafter we employ \textsc{numpy/scipy}
\citep{harris2020array,2020SciPyNMeth} together with \textsc{matplotlib
\citep{Hunter2007} }for post-processing and visualization.}
\end{figure*}

Figure \ref{fig1} shows the profiles of the large-scale flows, the
hydrodynamic $\alpha$ effect, {the convective turnover time,
$\tau_{c}$,} and the diffusivity profiles. We note the inverse sign
of the $\alpha$ effect tensor components and the rotational quenching
of the turbulent diffusivity profile toward the bottom of the convection
zone (marked by the dashed line). The profile of $\eta_{\mathcal{E}}$
remains unsaturated and it follows the parameters of the reference
convection zone model provided by MESA model.

\section{Results}

\subsection{Eigenvalue problem}

As the first step we consider the eigen value problem. It helps us
to define the critical thresholds of the dynamo instability and the
eigen modes profiles. In this case we simplify the model further and
neglect the overshoot region below the convection zone. At the bottom
of the convection zone, $r_{b}=0.728R$, we use the superconductor
boundary conditions for the mean electromotive force, $\mathcal{\overline{E}}_{\theta,\phi}=0$,
$\partial_{r}\mathcal{\overline{E}}_{r}=0$ and for the poloidal potential,
$A=0$. At the top, $r_{t}=0.99R$, we put $\partial_{r}\left(rA\right)=0$,
(radial magnetic field), $B=0$, $\mathcal{\overline{E}}_{r}=0$,
and $\partial_{r}\mathcal{\overline{E}}_{\theta}=0$. The numerical
integration in radius and latitude is done using the Galerkin method.
In the radial direction we decompose $A$, $B$ and $\overline{\boldsymbol{\mathcal{E}}}$
on the Chebyshev polynomials using the Gauss-Lobatto grid with 50
mesh points and in the latitudinal direction we use the associated
Legendre polynomials $P_{n}^{1}\left(\theta\right)$ and the Gauss-Legendre
grid with 72 points from pole to pole. To satisfy the radial boundary
conditions we use the basis recombination method \citep{Boyd2001}
. We put the \textsc{python} code for solution the eigen value problem
on \textsc{zenodo.}

\begin{figure*}
\includegraphics[width=0.9\textwidth]{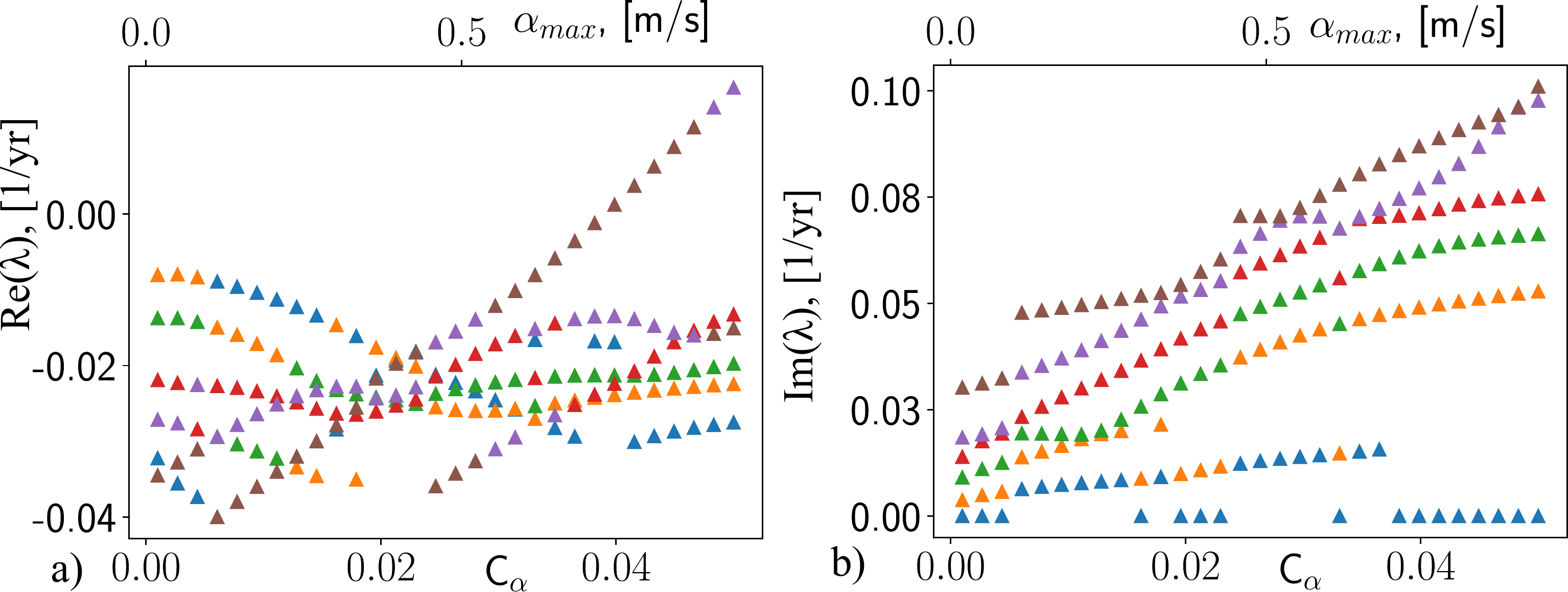}

\caption{\label{fig:local}a) Growth rates of the first six eigen odd dynamo
modes for the solar type dynamo model with the local mean electromotive
force, the x-axis show the maximum magnitude of the $\alpha_{\phi\phi}$
component in the convection zone; colors mark the different eigen
modes; b) shows the eigen frequency for each dynamo mode.}
\end{figure*}

\begin{table}
\caption{\label{tabe}Dependence of $C_{\alpha}^{(cr)}$ and the dynamo period
of the first unstable mode on parameters $\tau$ and $a_{E}$.}

\begin{tabular}{>{\raggedright}p{1cm}>{\raggedright}p{1.5cm}>{\raggedright}p{1.5cm}>{\raggedright}p{1cm}>{\raggedright}p{1cm}}
\hline 
$\tau$ & $a_{E}$ & $C_{\alpha}^{(cr)}$ & $\mathrm{P_{cyc}}$,

{[}yr{]} & Fig.\tabularnewline
\hline 
25d & 0.75 & 0.008 & 52 & \tabularnewline
10d & 0.75 & 0.015 & 28 & \tabularnewline
5d & 0.75 & 0.023 & 17 & \tabularnewline
1d & 0.75 & 0.033 & 10 & \tabularnewline
$\tau_{c}$(Fig\ref{fig1}c) & 0.25 & 0.025 & 19.8 & \ref{fig:nle}(a,b)\tabularnewline
$\tau_{c}$ & 0.5 & 0.015 & $\infty$ & \tabularnewline
$\tau_{c}$ & 0.75 & 0.01 & $\infty$ & \ref{fig:nle}(c,d)\tabularnewline
\end{tabular}
\end{table}
Figure \ref{fig:local} shows the growth rates and frequencies for
the first six eigen modes for the dynamo model with the local expression
of the mean electromotive force, i.e., $\overline{\boldsymbol{\mathcal{E}}}\equiv\overline{\boldsymbol{\mathcal{E}}}^{(0)}$.
Here, we discuss the odd parity modes, which are antisymmetric about
the equator. The property of the obtained dynamo modes is close to
results of \citet{Pipin2019c}. Yet, the full dynamo period is a bit
less than theirs. It is about 14 years. The difference is due to the
absence of the overshoot region in the given case. {The even
modes show very similar diagrams except the first instability threshold
for them is a bit higher than $C_{\alpha}^{\mathrm{cr}}\approx0.04$
for the odd modes. The same was found for the cases of the nonlocal
model considered below. }Typically, for this kind of dynamo models
the even modes have a higher instability threshold than the odd modes.
Also, we did not find them in the nonlinear runs. Therefore we omit
their discussion in below.

\begin{figure*}
\includegraphics[width=0.9\textwidth]{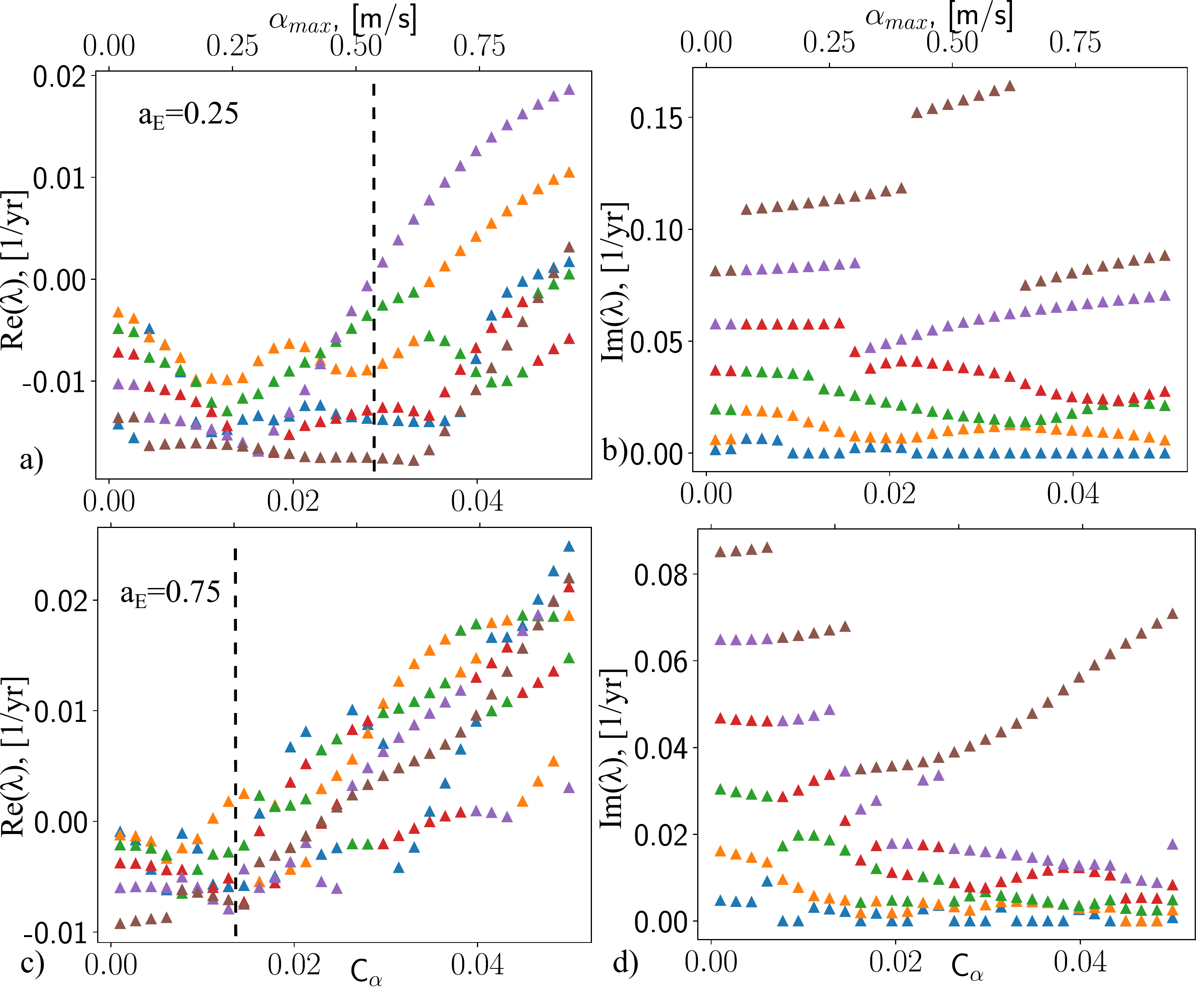} \caption{\label{fig:nle}The same as Fig.\ref{fig:local} for the nonlocal
model with the parameters $a_{E}=0.25$, top row, and $a_{E}=0.75$,
bottom row.}
\end{figure*}

Figure \ref{fig:nle} shows the instability diagram in the nonlocal
dynamo model for two cases $a_{E}=0.25$, and $a_{E}=0.75$. In these
solutions we put $\tau=\tau_{c}$ in the Eq(\ref{eq:nlc1}). {We
find that for $a_{E}\le0.1$ the eigen value instability diagram become
close to results of the model with the local $\boldsymbol{\overline{\mathcal{E}}}$.}
We see that in Fig.\ref{fig:nle} the instability threshold is lower
than for the case of the model with the local $\boldsymbol{\overline{\mathcal{E}}}$.
\citet{Rheinhardt2012} and \citet{Brandenburg2018a} found the same
tendency{. \citet{Rheinhardt2012} found that the cycle period
does not depend on $a_{E}$ and it seems to depend on $\tau$ only.
Indeed, if we put $\tau$ to be a constant, which is chosen from the
range of $\tau_{c}$ variations, we see that the instability threshold
and the dynamo period follows results of their paper (see Table\ref{tabe}).
Nevertheless, for the given $\tau=\tau_{c}$ profile the eigen dynamo
frequency show a dependence on $a_{E}$, at least for some of the
eigen modes. Figure \ref{fig:nle} shows a multiple instability with
several dynamo modes in proximity of the instability threshold of
the first mode, when the $\alpha$ effect parameter increases. Interesting
that within the given range of the $C_{\alpha}$ magnitudes, the growth
rates in the nonlocal dynamo are comparable with the local dynamo. }

\begin{figure}
\includegraphics[width=0.9\columnwidth]{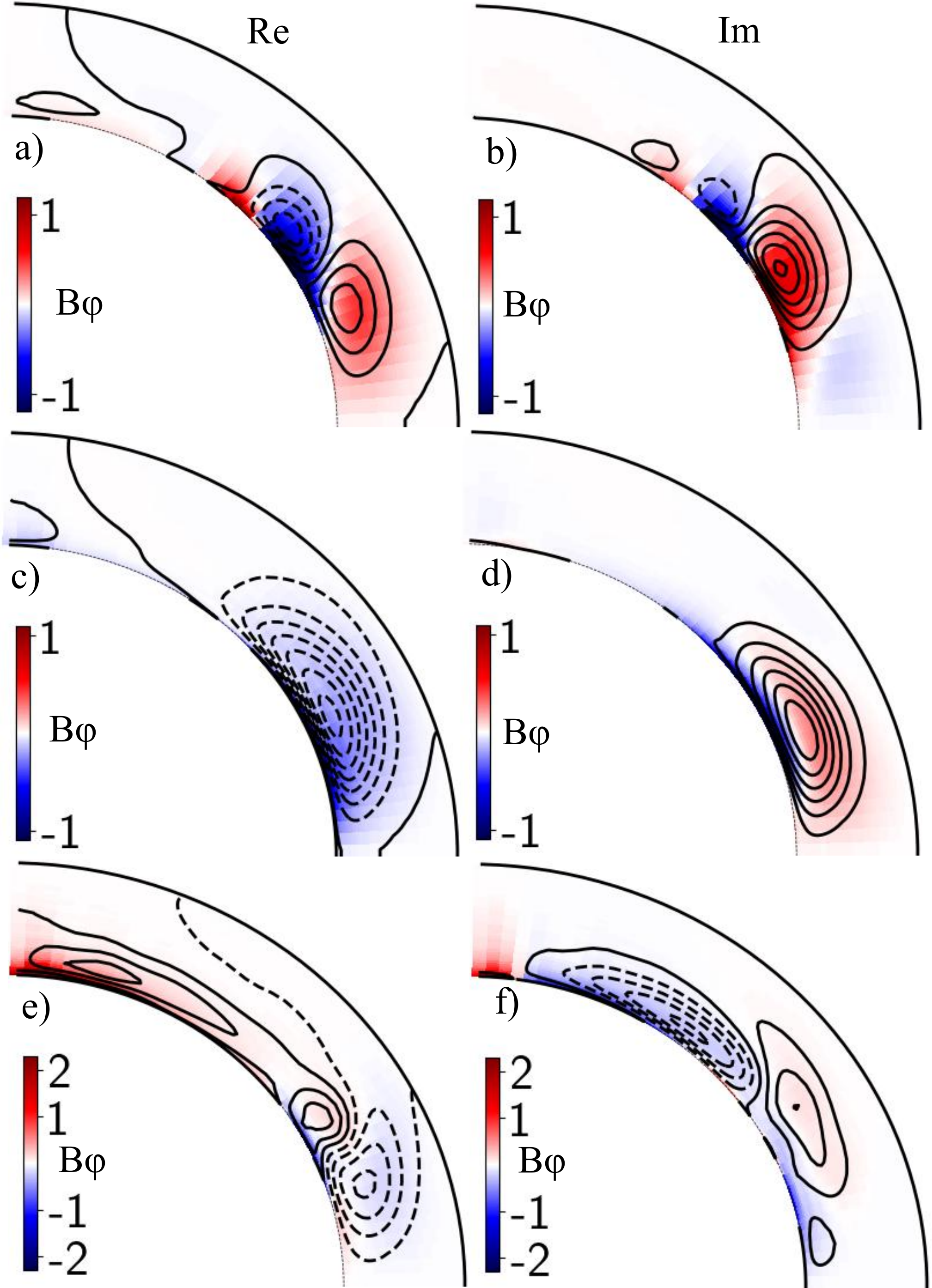} \caption{\label{fig:str}Spatial profiles of the real and imaginary parts of
the eigen solution, color shows the toroidal magnetic field and contours
show the streamlines of the poloidal field; we normalized the spatial
profiles to maximum value. {Panels a) and b) show the first
unstable mode for the local model; c) and d) show the same for the
local model with $a_{E}=0.25$; e) and f) show the second unstable
mode for $a_{E}=0.25$. }}
\end{figure}

{The spatial structure of the eigen modes show a difference.
It is illustrated in Fig.\ref{fig:str}. Here, we show the first unstable
mode of the model with the local $\boldsymbol{\overline{\mathcal{E}}}$,
as well. Comparing Fig.\ref{fig:str} a) and c), similar to \citet{Brandenburg2018a},
we see a smoothing effect of the turbulent diffusivity of $\boldsymbol{\overline{\mathcal{E}}}$
on the spatial variations scale of the toroidal magnetic in latitudinal
direction. Simultaneously we see a stronger concentration of the toroidal
magnetic field toward the bottom of the convection zone for the nonlocal
case. This can affect the increase of the dynamo period with the increase
of $a_{E}$, because of dynamo shifts to a place with the longer $\tau$
relaxation of $\boldsymbol{\overline{\mathcal{E}}}$. The first unstable
dynamo mode for $a_{E}=0.25$ describes the solar type dynamo waves
propagating toward the equator from bottom of the convection zone
with the period about 17 years. When we employ the solar profile $\tau=\tau_{c}$,
the solar type dynamo mode shows the lowest instability threshold
if $a_{E}<0.3$. The higher $a_{E}$ are possible in nonlinear supercritical
$C_{\alpha}$ cases. The dynamo period of this mode shows a decrease
with the increase of $a_{E}$. The second unstable mode at the diagram
of Figure \ref{fig:nle}a) has the dynamo period of about 64 years.
The dynamo period of this mode does not show dependence on parameter
$a_{E}$. It is likely because of its strong localization near the
bottom of the convection zone, see, Figs.\ref{fig:str} e) and f).}
We will look at it closely in the next subsection considering the
nonlinear solution. The primary feature of this mode is the enhance
of the toroidal magnetic field in the polar branch.

\begin{table*}
\caption{\label{tab2}The parameters of the nonlinear runs. Here, $B_{\phi}^{(\mathrm{max})}$
stands for the maximum of the toroidal magnetic field in the convection
zone; $B_{\mathrm{r}}^{(>60)}$ is the mean magnitude of the surface
radial magnetic field above 60$^{\circ}$ latitude; $F_{T}$ is the
total unsigned flux of the toroidal magnetic field in the convection
zone; $\mathrm{P_{cyc}}$ is the half period of the magnetic cycle.
We show the parameters of run M25 following results of \citet{Pipin21c},
noteworthy,$C_{\alpha}^{(cr)}=0.04$ for this case.}

\begin{tabular}{>{\raggedright}p{0.5cm}>{\raggedright}p{1.5cm}>{\raggedright}p{1.5cm}>{\raggedright}p{1.5cm}>{\raggedright}p{1.5cm}>{\raggedright}p{1cm}>{\raggedright}p{1.25cm}>{\raggedright}p{1cm}>{\raggedright}p{1cm}}
\hline 
 & Fig.  & $C_{\alpha}/C_{\alpha}^{(cr)}$  & Overshoot  & $a_{\mathcal{E}}$  & $B_{\phi}^{(\mathrm{max})}$,

{[}kG{]}  & $B_{\mathrm{r}}^{(>60)}$,

{[}G{]}  & $F_{T}$, $10^{24}${[}Mx{]}  & $\mathrm{P_{cyc}}$,

{[}yr{]}\tabularnewline
\hline 
N0  & \ref{tln0}, \ref{sn0},\ref{fl}  & 2.2  & +  & 0.5  & 3.2  & 6.5  & 1.  & 10.9\tabularnewline
N1  & \ref{fl}  & 1.1  & +  & 0.5  & 0.7  & 0.45  & $0.11$  & 30.2/25.1/272\tabularnewline
N2  & \ref{fl}  & 2.5  & +  & 1  & 3.5  & 8.9  & 1.5  & 11.8\tabularnewline
N3  & \ref{fl}  & 2.2  & -  & 0.5  & 3.1  & 3.5  & 0.9  & 9\tabularnewline
N4  & \ref{fl}  & 2.5  & -  & 1  & 4.1  & 5.7  & 1.2  & 9.3\tabularnewline
N5  & \ref{snn1},\ref{fl}  & 1.1  & +  & 0.75  & 1.1  & 0.53  & 0.13  & 28/35/326\tabularnewline
M25  & \ref{fl}  & 1.1  & +  & 0  & 2.5  & 5.6  & 1.1  & 10.6\tabularnewline
\end{tabular}
\end{table*}

\subsection{Nonlinear model}

The nonlinear dynamo model is based on the model of \citet{Pipin2019c}.
We consider the models both with include and exclude of the overshoot
region. Here, we discuss the kinematic dynamo models, i.e., we neglect
effect of the dynamo generated magnetic field on the heat transport
and the large-scale flow. In this case the dynamo saturation effects
are due to the magnetic helicity conservation, the ``algebraic''
quenching of $\alpha$ effect and the magnetic buoyancy. We can expect
that the last two effects are quenched in depth of the convection
zone because of the non-locality \citep{Brandenburg2018a}, which
is introduced by diffusivity of the mean electromotive force. The
parameters of the nonlinear run are listed in the Table\ref{tab2}.
To illustrate the nonlinear dynamo solution we choose the case $a_{E}=0.5$
($C_{\alpha}^{(cr)}=0.016$) and $a_{E}=1$ ($C_{\alpha}^{(cr)}=0.013$).

\begin{figure}
\includegraphics[width=0.99\columnwidth]{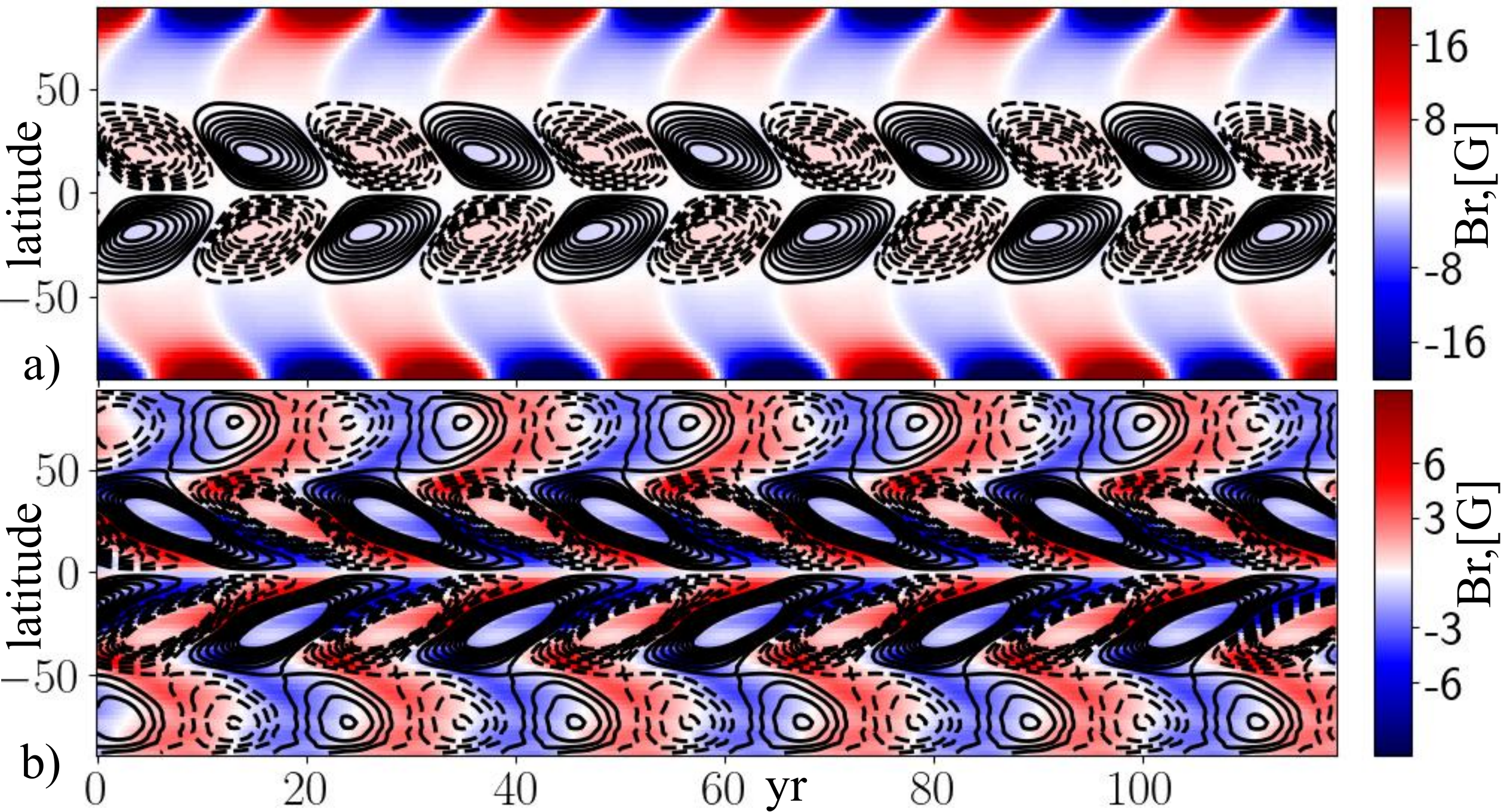}

\caption{\label{tln0}a) The time-latitude diagrams for the run N0, color image
shows the surface radial magnetic field and the toroidal magnetic
field at r=0.9R, is shown by contours in range $\pm1.$kG; b) the
same for the magnetic field in the overshoot layer, r=0.7R.}
\end{figure}

Our basic example is the case N0. The time-latitude diagrams of the
toroidal and radial magnetic field evolution are shown in Figure \ref{tln0}.
We see that the model with the nonlocal $\boldsymbol{\overline{\mathcal{E}}}$
preserves the basic properties of the earlier model of \citet{Pipin2019c}.
In the upper part of the convection zone the dynamo wave of the toroidal
magnetic field drifts toward the equator. Similar to the above cited
paper, this effect results from the joint action of the latitudinal
pumping, meridional circulation and the Parker-Yoshimura rule\citep{Yoshimura1975}.
Noteworthy, the magnitude of the $\alpha$ effect in run N0 is less
than the instability threshold of the reference case model with the
local $\boldsymbol{\overline{\mathcal{E}}}$. At the surface the radial
magnetic field drifts toward the poles at high latitudes and toward
the equator at low latitudes. The polarity sign of these branches
corresponds to the leading and following polarity of the sunspot activity.
Interesting that both the toroidal and radial magnetic field show
the extended 20 yrs branches of activity in overshoot region. This
can be important for the origin of the solar torsional oscillations.

\begin{figure}
\includegraphics[width=0.99\columnwidth]{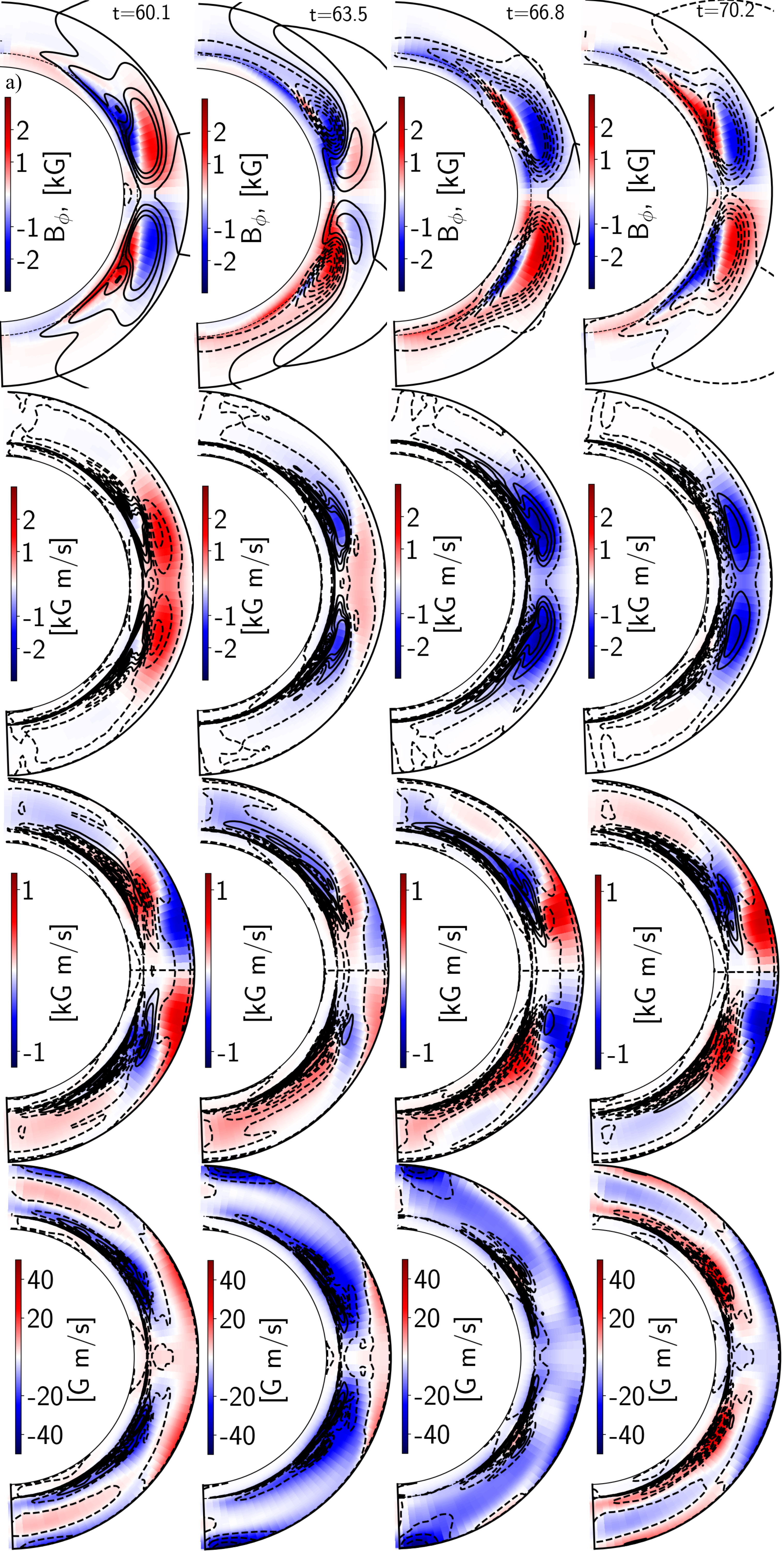}

\caption{\label{sn0} a) Snapshots for the magnetic field evolution in run
N0 for the half of the activity cycle, contours show streamlines of
the poloidal magnetic field; b) color image shows snapshots of $\overline{\mathcal{E}}_{r}$
, contours show the difference $\overline{\mathcal{E}}_{r}$-$\overline{\mathcal{E}}_{r}^{(0)}$
for the same range of value as the background color; c) and d) show
the same as b) for $\overline{\mathcal{E}}_{\theta}$ and $\overline{\mathcal{E}}_{\phi}$,
respectively.}
\end{figure}

Figure \ref{sn0} illustrates snapshots of the magnetic field and
mean electromotive force profiles in run N0 for the half of the magnetic
cycle. The magnetic activity shows the dynamo waves propagating from
the mid latitude at the bottom of the convection zone toward equator
at the surface. Evolution of the mean electromotive force shows the
qualitative similarity to results of the global convection simulations
of \citet{Racine2011}. Our model shows the two order magnitude less
$\overline{\mathcal{E}}_{\phi}$ than the results of the above cited
paper. This is because the $\alpha$ threshold for the dynamo instability
is order of magnitude less than the mixing length estimation of the
convection zone $\alpha$ (cf., Fig.\ref{fig1}b and Fig.\ref{fig:nle}).
Also, the magnitude of the dynamo generated magnetic field in our
model is less than in results of \citet{Racine2011}. The difference
$\overline{\boldsymbol{\mathcal{E}}}$-$\overline{\boldsymbol{\mathcal{E}}}^{(0)}$
shows the maximum near the bottom of the convection zone, in location
of the maximum of the toroidal magnetic field strength. This is because
of strong modulation of $\overline{\boldsymbol{\mathcal{E}}}^{(0)}$
by the dynamo generated magnetic field and the increase of $\eta_{\mathcal{E}}$
in the low part of the convection zone. 
\begin{figure}
\includegraphics[width=0.99\columnwidth]{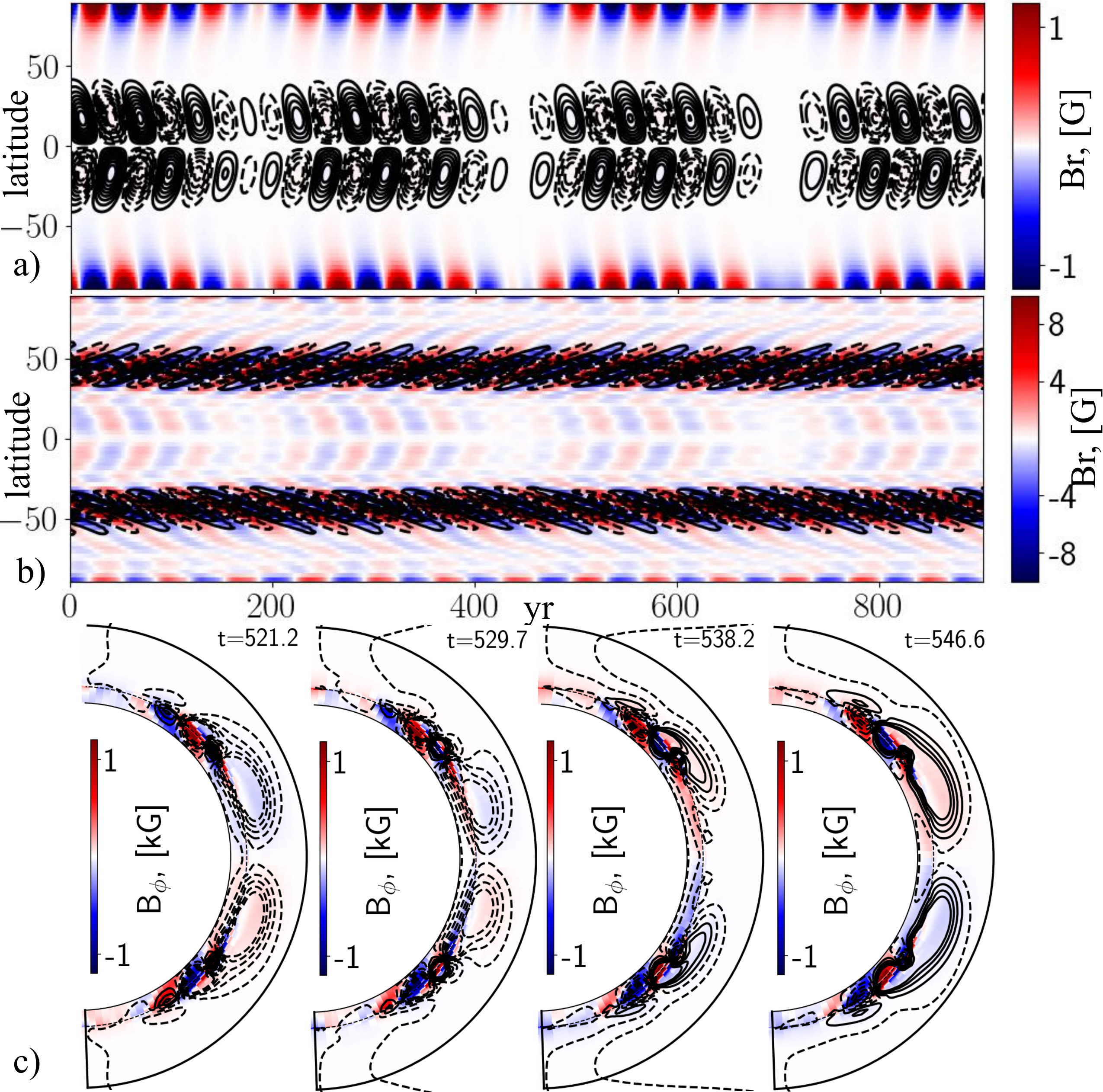}

\caption{\label{snn1}a) The time-latitude diagrams for the run N5, color image
shows the surface radial magnetic field and the toroidal magnetic
field at r=0.9R, is shown by contours in range $\pm50$G; b) the same
for the magnetic field in the overshoot layer, r=0.7R, contours of
the toroidal magnetic field are in the range $\pm1$kG; c) snapshots
of the magnetic field variations for the half of the magnetic cycle
of the equatorward propagating dynamo wave. }
\end{figure}

It is interesting to look at the dynamo solution in vicinity of the
instability threshold. The Fig.\ref{snn1} illustrates solution for
run N5, where we use a slightly overcritical $C_{\alpha}=1.1C_{\alpha}^{(cr)}$.
The solution show the order of magnitude less strength of the dynamo
generated magnetic field than in case N0. In the upper part of the
convection zone we find the qualitatively similar pattern of the magnetic
field oscillation with the full dynamo period about 60 years and the
Grand activity cycle of about 300 year period. The most of the magnetic
field flux is concentrated at the interface between the overshoot
region and the convection zone. There we see two different patterns
of the magnetic field oscillations. At low latitudes the dynamo waves
propagate toward equator with the period of magnetic activity about
30 years. At the mid latitudes there are waves with a slightly shorter
period of around 24 year. The beating of the two dynamo waves of the
different localization results to the long-term modulation of the
magnetic activity. {Both the solution of the eigen value problem
and the nonlinear runs shows that the dynamo period decreases with
the increase of the $\alpha$ effect magnitude \citep{Noyes1984}. }

\begin{figure}
\includegraphics[width=0.99\columnwidth]{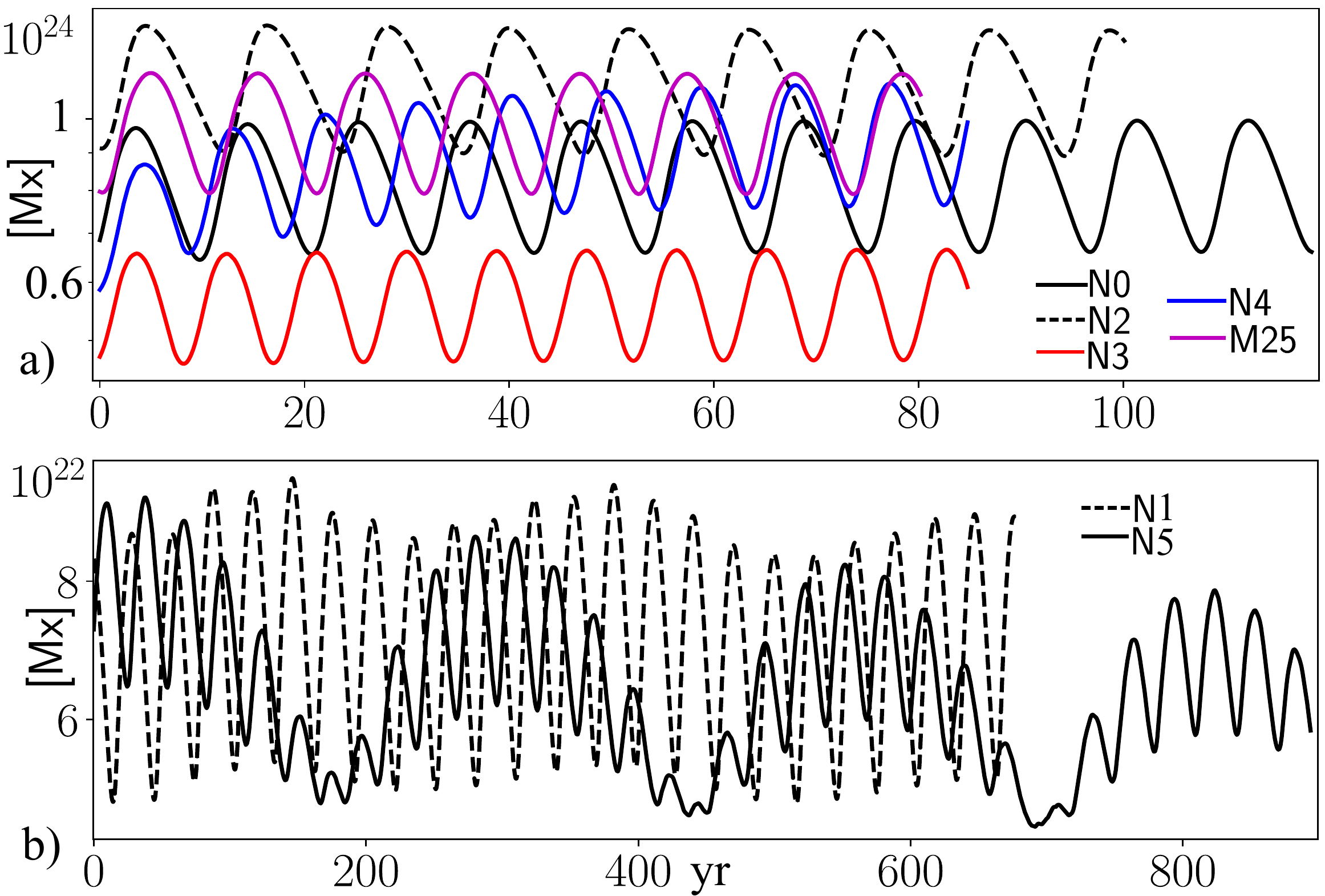}

\caption{\label{fl}The total unsigned flux of the toroidal magnetic field.
a) The runs with the single dynamo wave solution and supercritical
$C_{\alpha}\approx2.5C_{\alpha}^{(cr)}$; b) the same for the slightly
overcritical $C_{\alpha}\approx1.1C_{\alpha}^{(cr)}$ , and $a_{E}=0.5$
(N1), $a_{E}=0.75$ (N5).}
\end{figure}

The integral parameters of the runs are listed in the Table \ref{tab2}.
We find that increase of $a_{\mathcal{E}}$ increases the generated
magnetic flux. The models which include the overshoot layer show the
longer dynamo period than those confined to the convection zone. We
find that in nonlinear case an increase of $a_{E}$from $0.5$ to
1 does not result to a substantial increase of the dynamo period.
We take the run from \citet{Pipin21c} to compare the solar case dynamo
model with our runs. The run M25 (the above cited paper) employs the
slightly overcritical $C_{\alpha}=1.1C_{\alpha}^{(cr)}$, where, $C_{\alpha}^{(cr)}\approx0.04$,
and it has the higher amplitude of the $\alpha$ effect than the N's
runs presented here. The run M25 run shows the similar magnitude of
the total toroidal flux to the runs N0, N2 and N4. The difference
in the dynamo period between the runs is because of lower dynamo instability
threshold, $C_{\alpha}^{(cr)}$, for the nonlocal dynamo model. For
the slightly overcritical $C_{\alpha}$, the nonlocal dynamo can show
the long-term variations of the magnetic activity cycles if the parameter
$a_{E}$ is large enough, see Fig.\ref{snn1}a and Fig.\ref{fl}b
. These long-term variations are due to interference of the dynamo
modes of different spatial localization, see Fig.\ref{snn1}. The
run N5 shows two waves of magnetic activity with close periods of
28 and 35 years. These waves start from about 30$^{\circ}$ latitude
and propagate in opposite directions. The mode interaction, because
of the non-locality produce the long-term cycle of period 326 years.
Interesting that in the run N1 the long-term cycle disappear after
a while. In this case, the modes interaction seems to result into
nonlinear synchronization of two waves.

We find that the dipole type parity solution dominates in all the
runs. {The eigen value solution supports this conclusion. In
the nonlocal dynamo model smoothing of the mean electromotive force
by the magnetic diffusivity results into an increased concentration
of the toroidal field to the bottom of the convection zone.} The study
of \citet{Chatterjee2004} showed that this effects makes the dipole
type parity preferable. The question about dependence of this property
from the radial profiles of the turbulent parameters should be investigated
separately.

\section{Discussion}

We study effects of the nonlocal mean electromotive force in the solar
types dynamo models. Our formulation of the nonlocal $\overline{\boldsymbol{\mathcal{E}}}$
follows the approach suggested by \citet{Rheinhardt2012} and \citet{Brandenburg2018a}.
In following the results of the DNS, they suggested that the general
integro-differential equation for the mean electromotive force can
be replaced by the parabolic equation, see the Eqs(\ref{eq:nlc1},\ref{eq:nlc2}).
The temporal non-locality was suggested earlier by \citet{Brandenburg2003}
in discussion of the so called minimal $\tau$ approximation \citep{Brandenburg2005b}.
With this approach we formulate the dynamo model that steps over the
scale separation approximation. There are both the observational and
theoretical requirements for this step. In particular, both the mean-field
and flux-transport dynamo models can show a rather strong gradient
of the magnetic field near the boundaries of the dynamo domain. We
show some examples of this behavior for the distributed mean-field
dynamo model in the paper. Also, the solution of the flux transport
dynamo model shows several thin magneto shears of the different sign
in the close vicinity of the bottom of the convection zone \citep{Dikpati1999,Jouve2007,Kumar2019}.
{Noteworthy, that the width of the Lorentzians $\hat{\alpha}_{ij}$
and $\hat{\eta}_{ijk}$ can be different (see, \citealt{Brandenburg2008a}).
Therefore, our approximation for evolution equation of the mean electromotive
force in form of the parabolic equation Eq.(\ref{eq:nlc1}) requires
a further study.}

The study finds that the mean-field solar type dynamo models preserve
their basic properties even with the nonlocal formulation of the mean
electromotive force. Similar to \citet{Brandenburg2018a}, we find
that accounting for non-locality reduces the dynamo instability thresholds.
This effect results from the effective eddy diffusivity quenching
because of turbulent diffusion of the mean electromotive force. {The
increase of the the turbulent diffusivity of $\overline{\boldsymbol{\mathcal{E}}}$
results in a shifts of the maximum of dynamo wave toward the bottom
of convection zone. In this case the region of the dynamo instability
corresponds to a higher value of $\tau$ and a longer relaxation time
of $\overline{\boldsymbol{\mathcal{E}}}$. Following results of \citet{Rheinhardt2012}
this imply the longer dynamo periods, also, see Table \ref{tabe}.
}The effect increase the efficiency of the differential rotation and
the flux transport by the meridional circulation, as well. The diffusive
quenching of the turbulent electromotive force results to some other
interesting findings, as well. For example, we see that in the nonlocal
model, the growth rate of unstable modes is comparable to local cases
for the same magnitude of the $\alpha$ -effect. Reduction of the
dynamo instability growth rate is because of saturation of the turbulent
generation in the depth of the convection zone. Also, the diffusive
quenching of the magnetic buoyancy promotes the stronger amplitude
of the toroidal field in the nonlocal model in compare to the model
with local $\overline{\boldsymbol{\mathcal{E}}}$. It is interesting
to verify the nonlocal form of $\overline{\boldsymbol{\mathcal{E}}}$
for the $\alpha^{2}$ dynamo models. 

{The decrease of the dynamo instability threshold with the
increase of $a_{E}$ and the relatively low growth rates in vicinity
of $C_{\alpha}^{(cr)}$ promote generation of the several dynamo modes
simultaneously. Interesting that the solar type dynamo mode has the
lowest threshold only for $a_{E}<0.3$. Nevertheless, it can become
dominant for larger $a_{E}$ if the $C_{\alpha}$ is supercritical
enough. In this case the solar type mode wins in the nonlinear regime,
e.g., the runs N0, N2, N3 and N4. Results of \citet{Brandenburg2017A}
showed coexistence of two dynamo periods in activity of solar type
stars with period of rotation between 10 and 30 days. The theoretical
interpretation of this phenomenon is under debate. \citet{Pipin21c}
showed the doubling of the dynamo frequency for this interval of the
rotational periods. Whether several dynamo modes can coexist in the
supercritical dynamo regime for the fast rotating stars has to be
studied further. Here, we find the coexistence of two dynamo modes
in a marginal nonlinear regime. I this case, the magnetic field evolution
concentrates near the bottom of the convection zone and the different
dynamo modes show the different dynamo period and different localization.
Their interference result to the long-term variation solution if the
parameter $a_{E}$ is large enough and $C_{\alpha}$ is close $C_{\alpha}^{(cr)}$.
This phenomenon has the same nature as the long-term oscillations
because of the parity interaction of the two dynamo modes with close
frequency \citep{Ivanova1976,Brandenburg1989}. Here, we have the
dynamo modes of the same parity but the different localization and
different directions of the dynamo wave propagation (see, Fig\ref{snn1}).
The situation is completely different for the supercritical cases
when $C_{\alpha}>2C_{\alpha}^{(cr)}$ , where the only one dynamo
mode survives, while the dynamo instability analysis shows the number
of the unstable modes.}

Finally, the most important result of the paper is that the solar
type dynamo model survives in conditions of the nonlocal mean electromotive
force after relaxing the two scales separation approximation. Our
deal with the turbulent non-locality effects follows the hint of the
DNS \citep{Rheinhardt2012,Elstner2020G,Kandu2022B}. Despite the suggested
mean-field model lose the connection with the fundamental physical
laws, it demonstrates a reasonable way to study stellar dynamos beyond
the mean-field approximations limits.

{Data Availability.} The data underlying this article are available
by request. The python modules for solution of the eigen value problem
are on zenodo:(10.48550/arXiv.2302.11176).

{Acknowledgments} The author thanks the financial support of
the Ministry of Science and Higher Education of the Russian Federation
(Subsidy No.075-GZ/C3569/278)

 \bibliographystyle{mnras}
\bibliography{dyn}
 
\end{document}